\documentclass[aip, jcp, reprint, floatfix, 10pt]{revtex4-1}
\usepackage{amsmath}
\usepackage{amsfonts}
\usepackage{amssymb}
\usepackage{fancyvrb}
\usepackage{graphicx}
\usepackage{color}

\global\long\def\tr{\mathrm{tr}}

\global\long\def\im{\imath}

\graphicspath{ {figuresderiv/} }
\graphicspath{ {../figuresderiv/} }

\newcommand{\<} {\left\langle}
\renewcommand{\>} {\right\rangle}

\newcommand{\dg} {\dagger}
\newcommand{\pd} {{\phantom\dagger}}

\newcommand{\ci}[1] {c_{#1}^\pd}
\newcommand{\cid}[1] {c_{#1}^\dg}

\newcommand{\bG} {\mathbf{G}}
\newcommand{\bS} {\mathbf{\Sigma}}
\newcommand{\bGa} {\mathbf{\Gamma}}
\newcommand{\bU} {\mathcal{U}}

\newcommand{\qh}{\mathcal{H}}

\newcommand{\ql}{\mathcal{L}}
\newcommand{\qs}{\mathcal{S}}
\newcommand{\qr}{\mathcal{R}}
\newcommand{\qe}{\mathcal{E}}
\newcommand{\qi}{\mathcal{I}}

\let\Im\relax
\DeclareMathOperator{\Im}{Im}

\renewcommand{\subsection}[1] {}

\begin{document}

\title{Relaxation-limited electronic currents in extended reservoir simulations}

\author{Daniel Gruss}

\affiliation{Center for Nanoscale Science and Technology,
             National Institute of Standards and Technology,
             Gaithersburg, MD 20899, USA}
\affiliation{Maryland Nanocenter, University of Maryland,
             College Park, MD 20742, USA}

\author{Alex Smolyanitsky}

\affiliation{Applied Chemicals and Materials Division,
             National Institute of Standards and Technology,
             Boulder, CO 80305, USA}

\author{Michael Zwolak}

\email{mpz@nist.gov}

\affiliation{Center for Nanoscale Science and Technology,
             National Institute of Standards and Technology,
             Gaithersburg, MD 20899, USA}

\begin{abstract}
Open-system approaches are gaining traction in the simulation of charge transport in nanoscale and molecular electronic devices. In particular, ``extended reservoir'' simulations,  where explicit reservoir degrees of freedom are present, allow for the computation of both real-time and steady-state properties but require relaxation of the extended reservoirs. The strength of this relaxation, $\gamma$, influences the conductance, giving rise to a ``turnover'' behavior analogous to Kramers' turnover in chemical reaction rates. We derive explicit, general expressions for the weak and strong relaxation limits. For weak relaxation, the conductance increases linearly with $\gamma$ and every electronic state of the total explicit system contributes to the electronic current according to its ``reduced'' weight in the two extended reservoir regions. Essentially, this represents two conductors in series -- one at each interface with the implicit reservoirs that provide the relaxation. For strong relaxation, a ``dual'' expression -- one with the same functional form -- results, except now proportional to $1/\gamma$ and dependent on the system of interest's electronic states, reflecting that the strong relaxation is localizing electrons in the extended reservoirs. Higher order behavior (e.g., $\gamma^2$ or $1/\gamma^2$) can occur when there is a gap in the frequency spectrum. Moreover, inhomogeneity in the frequency spacing can give rise to a pseudo-plateau regime. These findings yield a physically motivated approach to diagnosing numerical simulations and understanding the influence of relaxation, and we examine their occurrence in both simple models and a realistic, fluctuating graphene nanoribbon.
\end{abstract}

\maketitle

\begin{figure}
\includegraphics[width=\linewidth]{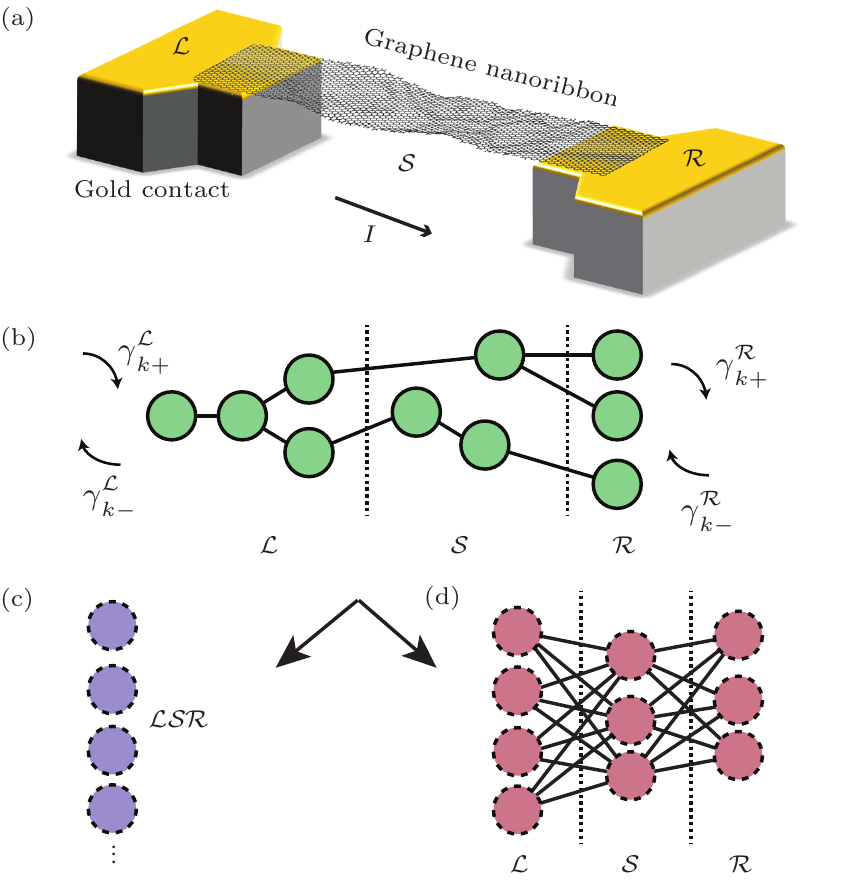}
\caption{
(a) A fluctuating graphene nanoribbon in aqueous solution between two gold substrates (water is not shown for clarity). (b) Schematic of the electronic modes of a $\ql\qs\qr$ system, where the couplings and onsite energies are arbitrary. Relaxation occurs only in the $\ql$ and $\qr$ regions, signified by the $\gamma_{k+}$ and $\gamma_{k-}$, where the superscript $\ql(\qr)$ is to explicitly indicate that the mode $k$ is in the left or right region at different chemical potentials. (c) For the small-$\gamma$ limit, the whole $\ql \qs \qr$ system is diagonalized to get a set of global modes. These determine the current, as the interface between $\ql \qs \qr$ and the implicit reservoirs is the rate limiting process. (d) For the large-$\gamma$ limit, each of the regions $\ql$, $\qs$, and $\qr$ are separately diagonalized, as the interface between $\ql (\qr)$ and $\qs$ determines the current.
\label{fig:diagram}}
\end{figure}

\begin{figure}
\includegraphics[width=\linewidth]{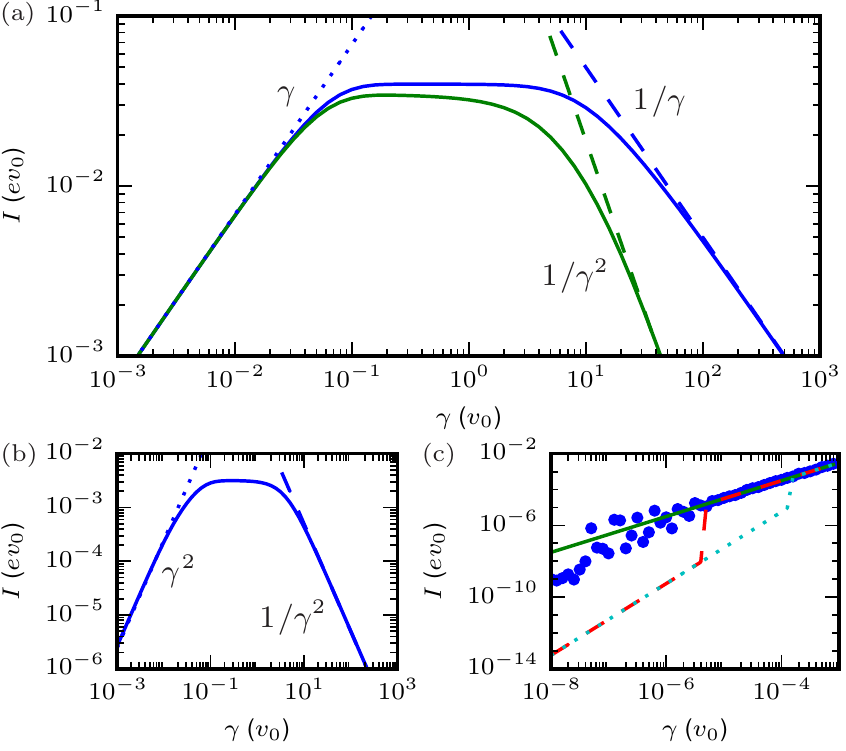}
\caption{
(a) The current, $I$, computed via the non-Markovian result, Eq.~\eqref{eq:totalcurr}, versus relaxation strength, $\gamma_k = \gamma$, for 1D reservoirs and $\qs$ being a single electronic mode either within (blue line) or outside (green line) the bias window. Here, the small-$\gamma$ behavior [blue dots, Eq.~\eqref{eq:smallgamma}] is nearly identical -- the shift in the energies of the $\ql\qs\qr$ system is negligible since the number of reservoir modes is large. For large $\gamma$, however, the behavior changes from $1/\gamma$ [blue dashes, Eq.~\eqref{eq:largegamma}] to $1/\gamma^2$ (green dashes, a fit) when the system's mode falls outside the bias window. (b) The current in the same 1D model with a bias such that no modes in $\ql\qs\qr$ or $\qs$ are in the bias window. The dotted and dashed blue lines show fits to $\gamma^2$ and $1/\gamma^2$, respectively. (c) Numerical integration errors for a single Lorentzian. In most cases, an insufficient error tolerance results in the integral behaving as $\gamma^2$ due to the tendency to exclude the bulk of the peak. Red dashed and cyan dotted lines show a Gaussian quadrature with two different error constraints. This systematic error is seen for most types of weighted integration. Hence, $\gamma^2$ behavior can be either real or due to numerical issues (easily identified/solved with smaller tolerances or by using methods with a predetermined error, such as Monte Carlo, shown as blue circles).
\label{fig:simplemod}}
\end{figure}


``Extended reservoir'' simulation recognizes that there is a hierarchy of length and time scales in transport. Particles (electrons, phonons, etc.) flow from very large reservoirs -- essentially, external sources or sinks -- into smaller, more confined regions before flowing through some ``system of interest'', a molecular junction, a nanotube, a graphene nanoribbon, etc. This concept gives rise to provably accurate simulation approaches that incorporate part of the reservoir explicitly into the simulation~\cite{velizhanin2015crossover2, Gruss2016, elenewski2017master}. For electron transport, this will give exactly the general results of Jauho, Meir, and Wingreen~\cite{meir1992landauer, jauho1994time, Haug-1996} for interacting and non-interacting systems alike (or, in the case of non-interacting systems, the Landauer formula~\cite{landauer1957spatial, di2008electrical}).

This approach requires the simulation of a larger system overall -- not only does one simulate the ``system of interest'', $\qs$, and possibly some small metallic leads, but rather that together with many extra degrees of freedom in the left and right ``extended reservoirs'', denoted $\ql$ and $\qr$ -- yet it yields a vast simplification for time-dependent phenomena. The two-time Green's functions can be replaced with a time-local (Markovian) master equation,
\begin{align} \label{eq:fullMaster}
\dot{\rho} = - \frac{\im}{\hbar} &[H_{\ql\qs\qr}, \rho]
    + \sum_{k \in \ql \qr} \gamma_{k+} \left( \cid{k} \rho \ci{k}
        - \frac{1}{2} \left \{ \ci{k} \cid{k}, \rho\right \}\right) \notag \\
    &+ \sum_{k \in \ql \qr} \gamma_{k-} \left( \ci{k} \rho \cid{k}
        - \frac{1}{2} \left \{ \cid{k} \ci{k}, \rho \right \} \right) ,
\end{align}
so long as the relaxation is weak enough (but---in order to extract properties of $\qs$---not too weak, i.e., not in the small-$\gamma$ regime we examine here~\cite{Gruss2016, elenewski2017master}). The density matrix, $\rho$, is for the $\ql\qs\qr$ system and $\cid{k} (\ci{k})$ are the creation (annihilation) operators for state $k$. The Hamiltonian is
\begin{equation} \label{eq:theHam}
H_{\ql\qs\qr} = H_\qs + H_\ql + H_\qr + H_\qi.
\end{equation}
where $H_\qs$ is for $\qs$ (including, potentially, many-body interactions), $H_{\ql(\qr)} = \sum_{k\in \ql(R)} \hbar \omega_k \cid{k} \ci{k}$ are for the ``extended reservoirs'', and $H_\qi = \sum_{k\in \ql\qr} \sum_{i\in \qs} (\hbar v_{ki} \cid{k} \ci{i} + \text{h.c.})$ is their interaction. The index $k$ includes all degrees of freedom (electronic state, spin, reservoir), while $\omega_k$ and $v_{ki}$ denote the level and hopping frequencies.

The first term in Eq.~\eqref{eq:fullMaster} describes Hamiltonian (unitary) dynamics of the $\ql\qs\qr$ system. The terms outside of the commutator reflect particle injection (depletion) into the state $k$ at a rate $\gamma_{k+}$ ($\gamma_{k-}$). These will relax the extended reservoirs to equilibrium -- a pseudo-equilibrium in terms of the isolated extended reservoir states~\cite{Gruss2016,elenewski2017master} -- if $H_\qi$ is removed when $\gamma_{k+} \equiv \gamma_k f^\alpha (\omega_k)$ and $\gamma_{k-} \equiv \gamma_k [1 - f^\alpha (\omega_k)]$, where $f^\alpha (\omega_k)$ is the Fermi-Dirac distribution in the $\alpha \in \{\ql,\qr\}$ reservoir.

Generically, one wants to simulate structures such as the one shown in Fig.~\ref{fig:diagram}(a), some molecule or nanoscale device (here, a graphene sensor) between two electronic reservoirs [see Fig.~\ref{fig:diagram}(b) for a schematic]. These types of structures are of interest to, e.g., sensing~\cite{willard2006directing, choi2012single, goldsmith2007conductance, goldsmith2008monitoring, sorgenfrei2011label, sorgenfrei2011debye} and sequencing~\cite{zwolak2005electronic, zwolak2008colloquium, lagerqvist2006fast, lagerqvist2007comment, lagerqvist2007influence, krems2009effect, chang2010electronic, TsutsuiTaniguchiYokotaEtAl2010, paulechka2016nucleobase, smolyanitsky2016mos2}. Within the extended reservoir approach, one models this setup by dividing the total system into three parts, the junction $\qs$ (molecule/structure, possibly including some part of the metallic leads) and the electronic reservoirs. The latter will be further split into the explicit degrees of freedom $\ql/\qr$ and implicit reservoirs that ensure proper sources/sinks are present. Considering one extended reservoir mode $k$, its relaxation -- its connection to the implicit reservoir $\qe_k$ -- is given by the Hamiltonian
\begin{equation} \label{eq:relaxation}
H_k = \hbar \omega_{k} \cid{k} \ci{k}
         + \sum_{\alpha \in \qe_k} \hbar \omega_{\alpha} \cid{\alpha} \ci{\alpha}
         + \sum_{\alpha \in \qe_k} \hbar t_{\alpha}
           \left( \cid{\alpha} \ci{k} + \cid{k} \ci{\alpha} \right) ,
\end{equation}
with $t_{\alpha}$ the coupling to the reservoir mode (which is related to the relaxation rate $\gamma$~\cite{Gruss2016, elenewski2017master}). The implicit reservoir is in equilibrium at some temperature and some chemical potential (which are different in the left and right regions), thus providing an infinite environment that will relax the mode $k$ to equilibrium in the absence of $\qs$. Including the environments $\qe_k$ gives a non-Markovian generalization of Eq.~\eqref{eq:fullMaster}. The Markovian approximation will be valid in the wide-band limit for $\qe_k$ and when $\gamma \ll k_B T/\hbar$, where $k_B$ is Boltzmann's constant and $T$ the temperature~\cite{Gruss2016, elenewski2017master}.

Both Eq.~\eqref{eq:fullMaster} and its full non-Markovian generalization have exact, closed form solutions for their steady-state currents, akin to the Meir-Wingreen formula~\cite{Gruss2016}. Our purpose here is to derive the small- and large-$\gamma$ limits of the steady-state for both the non-Markovian and Markovian cases. We deal solely with non-interacting electrons since this yields compact, illustrative expressions.


For the non-Markovian case, the total current is found by integrating out the environment in Eq.~\eqref{eq:relaxation}~\cite{jauho1994time, Gruss2016}, giving
\begin{align} \label{eq:totalcurr}
I = \frac{e}{2\pi}\int_{-\infty}^{\infty} d\omega \;
         &\left[f_\ql(\omega)-f_\qr(\omega)\right] \\
         &\times \tr \left[\bGa^{\ql}(\omega)\bG^{r}(\omega)
                           \bGa^{\qr}(\omega)\bG^{a}(\omega)\right] , \notag
\end{align}
where $f_{\ql(\qr)}$ is the equilibrium distribution in the left (right) reservoir, $\bGa^{\ql(\qr)}$ are the spectral densities, and $\bG^{r(a)}$ are the retarded (advanced) Green's functions (with bold capital letters indicating matrices). The environment being integrated out can either be the implicit reservoirs only or the implicit reservoirs plus $\ql$ and $\qr$. Both witl give the same result, but approximations will be easier using different forms in the different regimes, a fact that we will use below. The current, Eq.~\eqref{eq:totalcurr}, in the presence of relaxation has a turnover behavior~\cite{velizhanin2015crossover2, Gruss2016, elenewski2017master} $I_1 \,\, \left[ \mathrm{small} \, \gamma \right] \to I_2 \to I_3 \,\, \left[ \mathrm{large} \, \gamma \right]$,
where it first increases (regime 1), then plateaus (2), and then decreases (3).

\subsection{Weak Relaxation}

{\em Weak relaxation} -- When $\gamma$ is small, we will use the Green's functions for $\ql \qs \qr$ in Eq.~\eqref{eq:totalcurr}. For $H_{\ql}$ and $H_{\qr}$ separately diagonalized, the self-energies are
\begin{equation}
\bS^\ql_{ij} = - \frac{\im}{2} \sum_{k \in \ql}
                 \gamma_k \delta_{ij} \delta_{ik}
\qquad
\bS^\qr_{ij} = - \frac{\im}{2} \sum_{k \in \qr}
                 \gamma_k \delta_{ij} \delta_{ik} ,\label{eq:LSRSelf}
\end{equation}
where $\gamma$ is nonuniform~\cite{henderson2006determination, Gruss2016, zelovich2017parameter}. These are matrices on the whole $\ql \qs \qr$ system, but are zero when $i$ or $j$ are outside the respective reservoir region. The spectral and Green's functions are $\bGa^{\ql(\qr)} = -2 \Im \bS^{\ql(\qr)}$ and
\begin{equation}
\bG^r = \frac{1}{\omega - H_{\ql\qs\qr} - \bS^{\ql} - \bS^{\qr}} ,
\end{equation}
respectively. We rotate these expressions into the eigenbasis for $\ql\qs\qr$ through the unitary transform $\bU$, i.e.,
\begin{equation}
\left( \bU^\dagger H_{\ql\qs\qr} \bU \right)_{ij} = \tilde{\omega}_i \delta_{ij} ,
\end{equation}
where the $\tilde{\omega}_i$ are the eigenvalues of $H_{\ql \qs \qr}$. The transformation on the spectral functions and self-energies gives
\begin{equation}
\tilde{\bGa}^{\ql (\qr)} = \bU^\dagger \bGa^{\ql (\qr)} \bU ,
\qquad
\tilde{\bS}^{\ql (\qr)} = \bU^\dagger \bS^{\ql (\qr)} \bU .
\end{equation}
The quantity $\omega - H_{\ql\qs\qr}$ is now diagonal and the remaining terms in the denominator of $\bG^r$ are controlled by $\gamma$. The dominant terms are thus the diagonal components in this basis, as these diverge as $1/\gamma$, yielding
\begin{equation} \label{eq:AppG}
(\bG^r)_{ij} \approx \frac{1}{\omega - \tilde{\omega}_i - \tilde{\bS}^{\ql}_{ii} - \tilde{\bS}^{\qr}_{ii}} \delta_{ij} ,
\end{equation}
with $i,j \in \ql \qs \qr$. The off-diagonal components of the self-energies yield a higher order correction to this. This creates a sharply peaked Lorentzian to the lowest order, which can be integrated directly~\footnote{While the approximations can be done with the self-energies for both small and large $\gamma$, the entire integrand cannot be expanded due to the nature of the integral [Eq.~\eqref{eq:totalcurr}] over the Lorentzian. Some terms must be kept until after the integration is complete.}. Using Eq.~\eqref{eq:AppG} in Eq.~\eqref{eq:totalcurr} yields a double sum over eigenstates of $\ql \qs \qr$. The cross terms, $\int d\omega (\omega - \tilde{\omega}_i - \im \gamma / 2)^{-1} (\omega - \tilde{\omega}_j + \im \gamma / 2)^{-1} = (\pi \gamma^2 / 2)/(\gamma - \im \tilde{\omega}_i + \im \tilde{\omega}_j)$, in this sum behave as $\gamma^2$ when $\tilde{\omega}_i \neq \tilde{\omega}_j$ and $\gamma$ is small compared to the spacing $\left| \tilde{\omega}_i - \tilde{\omega}_j \right|$. In this limit, the total current, $I_1$, is
\begin{equation} \label{eq:smallgamma}
I_1 \approx \frac{e}{4} \sum_{i \in \ql\qs\qr}
            \frac{u_{i \ql}^2 u_{i \qr}^2}{u_{i \ql}^2 + u_{i \qr}^2}
            \left[ f_\ql (\tilde{\omega}_i) - f_\qr (\tilde{\omega}_i) \right]
    \propto \gamma.
\end{equation}
where
\begin{equation} \label{eq:matele}
 u_{i \ql (\qr)}^2 = \sum_{k \in \ql (\qr)} \gamma_k |\bU_{ik}|^2
\end{equation}
is the $\gamma$-weighted extent of the electronic modes in the extended reservoir regions. Equation~\eqref{eq:smallgamma} is a more general result than that derived in classical thermal transport to understand the influence of topological edge modes on the conductance \cite{chien2017thermal, chien2017topological}. This small-$\gamma$ limit is the same for the non-Markovian and Markovian [Eq.~\eqref{eq:fullMaster}] cases.

For low temperatures (on the electronic scale) and uniform reduced couplings (for $i$ in the bias window), Eq.~\eqref{eq:smallgamma} is proportional to the number of modes in the bias window. Thus, except for finite size effects [e.g., that change the mode structure and therefore change the form of the matrix elements in Eq.~\eqref{eq:matele}], the current for small $\gamma$ grows linearly with the number of modes in the extended reservoir. This means that the plateau region -- the flat region that follows the small-$\gamma$ regime, $I_2$ -- grows with the size of the extended reservoirs [and, indeed, this is why the {\em Markovian} approximation becomes accurate for simulating the Meir-Wingreen/Landauer conductance of $\qs$~\cite{Gruss2016, elenewski2017master}, i.e., the Markovian approximation is valid for $\gamma$ larger than that needed for the validity of Eq.~\eqref{eq:smallgamma}]. However, as we will see, inhomogeneity in the frequency spacing of $\ql \qs \qr$ modes can give rise to undesirable oscillatory features in between the small- and intermediate-$\gamma$ regimes (essentially, a symptom of finite sizes).

\subsection{Strong Relaxation}

{\em Strong relaxation} -- When $\gamma$ is large, it is advantageous to work with $\bG^r$ of $\qs$ only, giving the spectral functions
\begin{equation} \label{eq:spectral}
\bGa_{ij}^{\ql(\qr)}(\omega) = \sum_{k\in \ql(\qr)} v_{ik} v_{kj}
                               \frac{\gamma_k}{(\omega-\omega_k)^2+\gamma_k^2/4},
\end{equation}
where the $v$'s are the couplings in $H_\qi$. The highest order (in $1/\gamma$) term comes from the approximation
\begin{equation}
\bGa_{ij}^{\ql(\qr)}(\omega) \approx \sum_{k \in \ql(\qr)}
                                     \frac{4  v_{ik} v_{kj} }{\gamma_k}.
\end{equation}
That is, the relaxation must dominate all other energy scales. In practice, this means $\gamma_k \gg W$, where $W$ is the bandwidth of the extended reservoirs (e.g., all $| \omega_k |<W$), where we assume that the bias and temperature are small (so that the integration over $\omega$ is much smaller than $\gamma$~\footnote{This is necessary since the $\qe_k$ broaden the density of states and there is not a hard cutoff to the integration in Eq.~\eqref{eq:totalcurr}.}).

For just $\qs$, the Green's function is
\begin{equation}
\bG^r = \frac{1}{\omega - H_{\qs} - \bS^{\ql} - \bS^{\qr}} ,
\end{equation}
where $\bS^{\ql(\qr)}$ are not given by Eq.~\eqref{eq:LSRSelf}, but rather are the self-energies of $\ql(\qr)$ on $\qs$. Once again, we diagonalize the Hamiltonian component, i.e., $\omega-H_\qs$. The remaining terms are controlled by $1/\gamma$. As for small $\gamma$, the off-diagonal terms give a higher order contribution, hence
\begin{equation}
(\bG^r)_{ij} \approx \frac{1}{\omega - \omega_i -
                    \bS^{\ql}_{ii} - \bS^{\qr}_{ii}} \delta_{ij} ,
\end{equation}
where $\omega_i$ are the frequencies of $\qs$ in isolation (as compared to the entire $\ql\qs\qr$ system previously). This is again a Lorentzian, but now sharply peaked for large $\gamma$. The current, $I_3$, in the strong relaxation limit is then
\begin{equation} \label{eq:largegamma}
I_3 \approx \frac{e}{4} \sum_{i \in \qs} \frac{v_{i \ql}^2 v_{i \qr}^2} {v_{i \ql}^2 + v_{i \qr}^2}  \left[ f_\ql (\omega_i) - f_\qr (\omega_i) \right] \propto \frac{1}{\gamma},
\end{equation}
where
\begin{equation} \label{eq:coupleexpr}
v_{i \ql(\qr)}^2 = \sum_{k \in \ql (\qr)} \frac{4|v_{i k}|^2}{\gamma_k}.
\end{equation}
The current has the same form as the small-$\gamma$ limit except proportional to $1/\gamma$. As well, for small $\gamma$, the extent of the $\ql \qs \qr$ modes across a structure determines the current [Fig.~\ref{fig:diagram}(c)], whereas for large $\gamma$ the coupling between $\qs$ and $\ql(\qr)$ determines the current [Fig.~\ref{fig:diagram}(d)]. The renormalized form of the relevant coupling, Eq.~\eqref{eq:coupleexpr}, is a reflection of the Zeno effect~\cite{itano1990quantum}, where electrons are attempting to hop out of the reservoir into the system, but they are being strongly ``measured'' by the relaxation.

The Markovian case, though, is different than Eq.~\eqref{eq:largegamma},
\begin{equation} \label{eq:largegammamark}
\bar{I}_3 \approx \frac{e}{4} \sum_{i \in \qs}
                  \frac{\bar{v}_{i \ql}^2 \bar{v}_{i \qr}^2}
                       {\bar{v}_{i \ql}^2 + \bar{v}_{i \qr}^2}
          \propto \frac{1}{\gamma},
\end{equation}
where
\begin{equation} \label{eq:coupleexprmark}
\bar{v}_{i \ql(\qr)}^2 = \sum_{k \in \ql (\qr)} \frac{4 |v_{i k}|^2}{\gamma_k}  f_{\ql(\qr)}(\omega_k).
\end{equation}
This is due to the fact that the Markovian equation of motion populates the reservoir modes according to their isolated frequencies, i.e., it is a pseudo-equilibrium that does not account for the broadening of the density of states. Indeed, the Markovian approximation is not valid in this limit for this very reason~\cite{Gruss2016, elenewski2017master}.


\begin{figure}
\includegraphics[width=\linewidth]{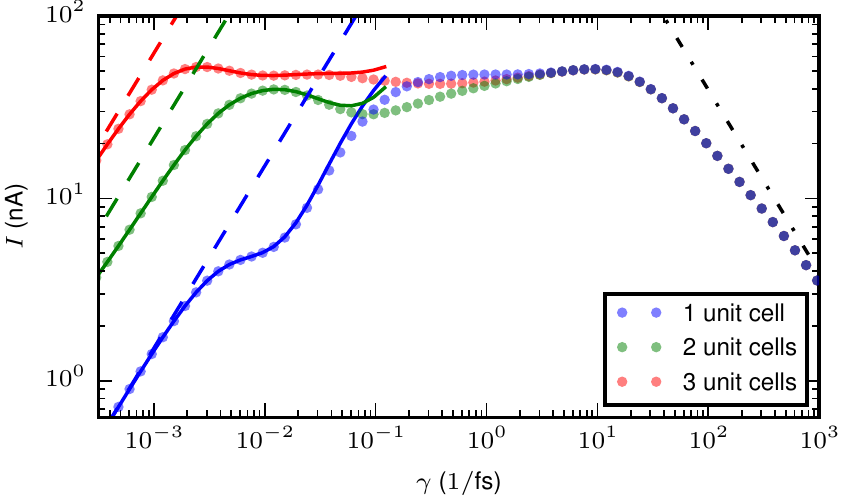}
\caption{
The electronic current, $I$ [from Eq.~\eqref{eq:totalcurr}], through a graphene nanoribbon suspended in aqueous solution between two gold substrates, plotted with an increasing number of Au unit cells (i.e., the contact between the gold and the graphene is kept fixed but the reservoir regions are made deeper). The dashed lines are the exact results for the small-$\gamma$ limit [Eq.~\eqref{eq:smallgamma}], which generally increase with the additional number of gold layers. However, within this small-$\gamma$ regime, finite-size effects prevent a simple linear scaling of the current with the extended reservoir size, i.e., the total current is not dependent on solely the size of the extended reservoir but also on the structure of the electronic states. All cases go to the same large-$\gamma$ limit (dash-dot line) since $\qs$ is identical. In the intermediate regime, a larger plateau forms for the larger extended reservoirs. However, inhomogeneous mode spacing will create inhomogeneity in the turnover points, which in turn will create features in the small- to intermediate-$\gamma$ region. That is, even outside the formally small-$\gamma$ region, the results reflect the distribution in important quantities in the small-$\gamma$ regime (solid lines show the contribution of the modes in and around the bias window in the $\ql\qs\qr$ basis).
\label{fig:graphenelims}}
\end{figure}

Equations~\eqref{eq:smallgamma},~\eqref{eq:largegamma}, and~\eqref{eq:largegammamark} are our main results. Figure~\ref{fig:simplemod}(a) shows the current, $I$, versus $\gamma$ for a simple example: A system with only a single mode symmetrically coupled to identically-sized linear (1D) reservoirs in $\ql$ and $\qr$ all with strength $v_0$. When the system mode has zero onsite energy, the small- and large-$\gamma$ limits display $\gamma$ and $1/\gamma$ behavior, respectively. However, if the system mode is shifted outside of the bias window, the leading term for large $\gamma$ becomes $1/\gamma^2$. Similarly, Fig.~\ref{fig:simplemod}(b) shows a total $\ql\qs\qr$ system that has no modes in the bias window, giving $\gamma^2$ dependence for small $\gamma$. The error inherent in numerical integration of a Lorentzian can also yield incorrect limits for small $\gamma$ [Fig.~\ref{fig:simplemod}(c)].

We also apply this method to a suspended graphene nanoribbon [Fig.~\ref{fig:diagram}(a)]. In this example, $\ql$ and $\qr$ are the gold contacts and $\qs$ is the ribbon itself. The graphene Hamiltonian is
\begin{equation} \label{eq:exphop}
\qh = \sum_{\< i j \>} v_0 e^{-l/\lambda} \cid{i} \ci{j} ,
\end{equation}
where $v_0 \approx-3$~eV is the hopping energy between $p_z$ orbitals in pristine graphene, $l$ is the carbon-carbon bond stretching, and $\lambda \approx 0.047$~nm (parameters are a rearrangement of those in Ref.~\onlinecite{cosma2014strain}). For the substrate, the gold atoms have a $-3$~eV hopping energy and an onsite energy of $-1.6$~eV, representing the $6s$-conduction band~\cite{wahiduzzaman2013dftb}. The carbon-gold coupling is of similar form to Eq.~\eqref{eq:exphop}, with the same $v_0$ and $\lambda \approx 0.11$~nm.

Figure~\ref{fig:graphenelims} shows the current as a function of $\gamma$ for a $50$~meV bias and $200$~meV Fermi level. The small-$\gamma$ prefactor generally increases with reservoir size, but the exact form depends on the $\ql \qs \qr$ mode structure. For the smallest reservoir size (blue data points), there are only four $\ql \qs \qr$ modes in the bias window. These ``turnover'' out of the small-$\gamma$ regime at about $\gamma \approx 3 \times 10^{-3}$ fs$^{-1}$. Almost simultaneously, a $\gamma^2$ contribution turns on from a mode just outside the bias window, giving the quasi-linear regime before the plateau. This can be made concrete by defining projection operators onto the subspace of these five modes (or even separate projectors onto the bias window modes and the one just outside the bias window). The blue solid line shows the contribution by only these handful of modes, clearly showing that they give the features in between the small- and intermediate-$\gamma$ regimes. In addition, the defining role of $\ql \qs \qr$ modes in the transition to the plateau regime allows one to determine when the mode spectrum (i.e., coupling across the whole structure versus frequency) is uniformly converging or still displaying finite-size effects. The green and red data points have a more rapid rise to the plateau (due to the larger number of $\ql \qs \qr$ modes), but inhomogeneity of the mode spacings gives rise to a pseudo-plateau region, where oscillations occur before transitioning into the real plateau.


For both weak and strong relaxation, the current depends on the amplitude of the eigenmodes at the boundaries, but \emph{different} boundaries and therefore different eigenmodes. Going from small to large $\gamma$ changes the relevant contact resistance from that between $\ql\qs\qr$ and the implicit reservoirs to that between $\qs$ and $\ql$ and $\qr$. However, the $\ql\qs\qr$ modes -- and hence their spacing and uniform convergence -- play the dominant role in the transition from the small-$\gamma$ to the plateau regime, where the current reflects the intrinsic conductance of $\qs$~\cite{Gruss2016, elenewski2017master}. We note, of course, that the system of interest does not have a unique conductance, independent of how it is contacted. Thus, the intrinsic conductance reflects the natural conductance of the setup with all $\gamma$ dependence eliminated.

Open-system simulations of the type we address here are increasingly in use for simulating nanoelectronics~\cite{sanchez2006molecular, subotnik2009nonequilibrium, ajisaka2012nonequlibrium, hod2016driven, morzan2017electron, zelovich2017parameter}. These also have a relation to closed system simulations of real-time dynamics~\cite{di2004transport, stefanucci2004time, bushong2005approach} (including of particle transport in cold-atom setups~\cite{chien2012bosonic, chien2013interaction, chien2014landauer, gruss2016energy}). To successfully employ this simulation approach, one should be in between the two limits and, in the case of Eq.~\eqref{eq:fullMaster}, with $\gamma$ still weak enough that a Markovian approximation is valid~\cite{Gruss2016, elenewski2017master}. The compact analytic forms we derive allow one to benchmark these simulations, as well as make simple predictions for other scenarios (e.g., quantum dot systems or molecules with variable contacts) where the relaxation/tunneling rate can be tuned. Moreover, the physics associated with this whole turnover process is analogous to Kramers' turnover in chemical relaxation rates~\cite{velizhanin2015crossover2, Gruss2016}, thus giving a physically intuitive conceptual paradigm for simulating various transport processes, from thermal/energy to electronic.


Daniel Gruss acknowledges support under the Cooperative Research Agreement between the University of Maryland and the National Institute of Standards and Technology Center for Nanoscale Science and Technology, Award 70NANB14H209, through the University of Maryland. Alex Smolyanitsky gratefully acknowledges support from the Materials Genome Initiative.

\end{document}